\RequirePackage{lineno}
\documentclass[prd,twocolumn,showpacs,amsmath,amssymb]{revtex4}
\usepackage{graphicx}
\usepackage{dcolumn}
\usepackage{bm}
\usepackage{epsfig}
\usepackage{overpic}
\usepackage{verbatim}
\newsavebox{\tablebox}
\usepackage[colorlinks,linkcolor=blue,anchorcolor=red,citecolor=blue]{hyperref}

\usepackage{rotating}

\newcommand {\NIM}    {Nucl.{} Instrum.{} Meth.{} }

\newcommand {\PR}     {Phys.{} Rev.{} }

\newcommand{\ee}{e^+e^-}
\newcommand{\Rmnum}[1]{\uppercase\expandafter{\romannumeral #1}}
\def\jpsigaminv{J/\psi\to\gamma + ~\rm {invisible}}
\def\pipijpsi{\psip\to\pp\jpsi}
\newcommand{\BR}{{\cal B}}

\newcommand{\psip}{\psi(3686)}

\newcommand{\jpsi}{J/\psi}

\def\gevcc{\ifmmode {\mathrm{\ Ge\kern -0.1em V}/c^2}\else
                   {\textrm{Ge\kern -0.1em V}/$c^2$}\fi}%
\def\gev{\ifmmode {\mathrm{\ Ge\kern -0.1em V}}\else
                   {\textrm{Ge\kern -0.1em V}}\fi}%
\def\mevcc{\ifmmode {\mathrm{\ Me\kern -0.1em V}}\else
                   {\textrm{Me\kern -0.1em V}}\fi}%

\newcommand{\Mrecpipi}{M^{rec}_{\pi^+\pi^-}}
\newcommand{\Egam}{E_\gamma^*}
\newcommand{\pp}{\pi^+\pi^-}

\newcommand{\bfg}{\begin{figure}}
\newcommand{\efg}{\end{figure}}
\newcommand{\bitm}{\begin{itemize}}
\newcommand{\eitm}{\end{itemize}}
\newcommand{\bnum}{\begin{enumerate}}
\newcommand{\enum}{\end{enumerate}}
\newcommand{\btbl}{\begin{table}}
\newcommand{\etbl}{\end{table}}
\newcommand{\btbu}{\begin{tabular}}
\newcommand{\etbu}{\end{tabular}}

\newcommand{\beq}{\begin{equation}}
\newcommand{\edq}{\end{equation}}

\begin{document}
\normalsize
\parskip=5pt plus 1pt minus 1pt

\title{\boldmath Search for the decay $\jpsigaminv$}
\author{\small
M.~Ablikim$^{1}$, M.~N.~Achasov$^{10,c}$, P.~Adlarson$^{64}$, S. ~Ahmed$^{15}$, M.~Albrecht$^{4}$, A.~Amoroso$^{63A,63C}$, Q.~An$^{60,48}$, ~Anita$^{21}$, Y.~Bai$^{47}$, O.~Bakina$^{29}$
% ...
\\
\vspace{0.2cm}
(BESIII Collaboration)\\
\vspace{0.2cm} {\it
$^{1}$ Institute of High Energy Physics, Beijing 100049, People's Republic of China\\
% ...
$^{4}$ Bochum Ruhr-University, D-44780 Bochum, Germany\\
% ...
$^{10}$ G.I. Budker Institute of Nuclear Physics SB RAS (BINP), Novosibirsk 630090, Russia\\
% ...
$^{15}$ Helmholtz Institute Mainz, Johann-Joachim-Becher-Weg 45, D-55099 Mainz, Germany\\
% ...
$^{21}$ Indian Institute of Technology Madras, Chennai 600036, India\\
% ...
$^{29}$ Joint Institute for Nuclear Research, 141980 Dubna, Moscow region, Russia\\
% ...
$^{47}$ Southeast University, Nanjing 211100, People's Republic of China\\
$^{48}$ State Key Laboratory of Particle Detection and Electronics, Beijing 100049, Hefei 230026, People's Republic of China\\
% ...
$^{60}$ University of Science and Technology of China, Hefei 230026, People's Republic of China\\
% ...
$^{63}$ (A)University of Turin, I-10125, Turin, Italy; (B)University of Eastern Piedmont, I-15121, Alessandria, Italy; (C)INFN, I-10125, Turin, Italy\\
$^{64}$ Uppsala University, Box 516, SE-75120 Uppsala, Sweden\\
% ...
$^{c}$ Also at the Novosibirsk State University, Novosibirsk, 630090, Russia\\
% ...
}
}

\date{\today}

\begin{abstract}

We search for $\jpsi$ radiative decays into a weakly interacting neutral particle, namely an invisible particle, 
using the $\jpsi$ produced through the process $\pipijpsi$ in a 
data sample of $(448.1\pm2.9)\times 10^6$ $\psip$ decays
collected by the BESIII detector at BEPCII. 
No significant signal is observed.
Using a modified frequentist method, upper limits on the branching fractions are set under different assumptions of invisible particle masses up to 1.2~$\gevcc$. 
The upper limit corresponding to an invisible particle with zero mass is 7.0$\times 10^{-7}$ at the 90\% confidence level.

\end{abstract}


\maketitle

\section{INTRODUCTION}

Understanding the nature of dark matter and finding direct evidence for its existence are among the primary goals of contemporary astronomy and particle physics~\cite{Bertone:2004pz, ArkaniHamed:2008qn}.
Numerous experiments aim for the direct detection of dark matter, but no solid evidence has yet been found~\cite{PDGweb,Ambrosi:2017wek,Cui:2017nnn,Akerib:2016vxi,Aprile:2017iyp}.
A series of Supersymmetric Standard Models~\cite{Fayet:1977yc}, including the Next-to-Minimal Supersymmetric Model (NMSSM)~\cite{Ellwanger:2009dp,Maniatis:2009re}, 
predict a light CP-odd pseudoscalar Higgs boson $A^0$ and a series of neutralinos.
The light stable neutralino~($\chi^0$), in particular, which is one possible explanation for the 511 keV $\gamma$ ray feature observed by the INTEGRAL satellite~\cite{Jean:2003ci},
is one of the candidates for dark matter particles~\cite{Shrock:1982kd,Gunion:2005rw}.
The $\chi^0$ can couple with Standard Model particles via the $A^0$ boson,
and the $A^0$ can be produced in the radiative decay of a quarkonium vector state, $V$~\cite{Fayet:1981rp,Wilczek:1977zn,Fayet:2007ua}.
The branching ratio of such a radiative decay is:

\beq
\label{eq:Theory}
\frac{\mathcal{B}(V \to \gamma A^0)}{\mathcal{B}(V \to \mu^+\mu^-)}=\frac{G_Fm_q^2g_q^2C_{QCD}}{\sqrt{2}\pi\alpha}\bigg(1-\frac{m_{A^0}^2}{m_{V}^2}\bigg),
\edq
where $m_{A^0}$, $m_{V}$ and $m_q$ are the masses of the $A^0$, the quarkonium state, and the corresponding quark, respectively;
$\alpha$ is the fine structure constant;
$G_F$ is the Fermi coupling constant;
$C_{QCD}$ is the combined QCD radiative and relativistic corrections~\cite{Nason:1986tr}, which depends on $m_{A^0}$;
and $g_q$ is the Yukawa coupling of the $A^0$ field to the quark-pair, and is $g_c = \rm{cos}\theta_A/\tan\beta$ for the charm quark and  $g_b = \rm{cos}\theta_A\tan\beta$ for the bottom quark,
where $\tan\beta$ is the usual ratio of vacuum expectation values
and $\theta_A$ is the Higgs mixing angle~\cite{Gunion:2005rw}.

The CLEO-c~\cite{Insler:2010jw}, BaBar~\cite{Aubert:2008as,delAmoSanchez:2010ac} and  Belle~\cite{Seong:2018gut} experiments have performed similar searches for $\jpsi$ or $\Upsilon$ radiative decays into invisible particles, and no signal was observed.
The upper limits at the 90\% confidence level (C.L.) for the branching fraction of the decay $\jpsigaminv$, $\mathcal{B}(\jpsigaminv)$, are in the range $(2.5\sim 6.3)\times10^{-6}$, depending on the mass of $A^0$~\cite{Insler:2010jw}, where $\mathcal{B}(\jpsigaminv)$ is the product of $\mathcal{B}(\jpsi\to\gamma + A^0)$ and $\mathcal{B}(A^0\to\chi^0\bar{\chi}^0)$.
It is worth noting that the decay process $\jpsi\to\gamma\nu\bar{\nu}$, which is allowed in the Standard Model, is an irreducible background in this analysis, but the predicted branching fraction is only $0.7\times10^{-10}$, which is far below our experimental sensitivity~\cite{Gao:2014yga}. 
Thus, this background is neglected.

In this paper, we search for the $\jpsigaminv$ decay using $\jpsi$ produced through the process $\pipijpsi$  in a data sample of $(448.1\pm2.9)\times 10^6$ $\psip$ decays collected with the BES\Rmnum{3} detector.

\section{BES\Rmnum{3} DETECTOR AND MONTE CARLO SIMULATION}
\label{detmc}

The BESIII detector is a magnetic
spectrometer~\cite{Ablikim:2009aa} located at the Beijing Electron
Positron Collider (BEPCII)~\cite{bepcii}. The
cylindrical core of the BESIII detector consists of a helium-based
 multilayer drift chamber (MDC), a plastic scintillator time-of-flight
system (TOF), and a CsI(Tl) electromagnetic calorimeter (EMC),
which are all enclosed in a superconducting solenoidal magnet
providing a 1.0~T magnetic field. The solenoid is supported by an
octagonal flux-return yoke with resistive plate counter muon
identifier modules interleaved with steel. The acceptance of
charged particles and photons is 93\% over $4\pi$ solid angle. The
charged-particle momentum resolution at $1~{\rm GeV}/c$ is
$0.5\%$, and the $dE/dx$ resolution is $6\%$ for the electrons
from Bhabha scattering. The EMC measures photon energies with a
resolution of $2.5\%$ ($5\%$) at $1$~GeV in the barrel (end cap)
region. The time resolution of the TOF barrel part is 68~ps, while
that of the end cap part is 110~ps. 

The performance of the BESIII detector is evaluated using a {\sc geant4}-based~\cite{geant4}  Monte Carlo (MC) program that includes the description of the detector geometry and response. 
To check for potential backgrounds, an inclusive MC sample of $\psip$ decays is used.  The sample includes approximately the same number of $\psip$ decays as in data.
The production of the $\psip$ resonance is simulated by the MC event generator {\sc kkmc}~\cite{kkmc}, taking into account the beam energy spread; the known decay modes are generated using {\sc evtgen}~\cite{evtgen} with the branching fractions as given by the particle data group (PDG)~\cite{PDGweb}; the unknown decay modes are modeled with the {\sc lundcharm} model~\cite{lundcharm}.
Signal MC samples, corresponding to $\pipijpsi$ with the subsequent decay $\jpsigaminv$, are used to evaluate the detection efficiencies and model the line shapes of variables of interest.  The samples are generated under different assumptions for $m_{A^0}$.
In these signal MC samples, the decay $\jpsigaminv$ is modeled with an angular distribution of $1 + \cos^2 \theta_\gamma$ ($\theta_\gamma$ is the angle of the radiative photon relative to the positron beam direction in the $\jpsi$ rest frame).
Throughout the text, the decay $\pipijpsi$ is modeled according to the formulas and measurement in Ref.~\cite{Bai:1999mj}.
In this analysis, detailed MC studies indicate that the dominant backgrounds are from $\pipijpsi$ with subsequent decays $\jpsi\to\gamma\pi^0$, $\gamma\eta$ and $\gamma K_L K_L$. 
These backgrounds are each generated exclusively with more than 100 times the statistics in data, where the decays of $\jpsi\to\gamma \pi^0$ and $\gamma \eta$ are generated with the angular distribution of $1 + \cos^2 \theta_\gamma$, and $\jpsi\to\gamma K_L K_L$ is modeled  with the partial wave analysis (PWA) results of $\jpsi\to\gamma K_S K_S$~\cite{gammaksks} by assuming isospin symmetry. 
Many potential backgrounds of the form $\pipijpsi$ with $\jpsi$ decaying into purely neutral particles in the final states, or with large branching fractions, are generated exclusively with different generators, $i.e.$ $\jpsi\to\gamma\eta'$, $\gamma\eta(1405)$ and $\gamma\eta_c$ with the angular distribution of $1 + \cos^2 \theta_\gamma$; $\jpsi\to\gamma\pi^0\pi^0$ and $\gamma\pi^+\pi^-$ according to PWA results of $\jpsi\to\gamma \pi^0\pi^0$~\cite{gampi0pi0} with isospin symmetry assumption; $\jpsi\to\gamma K^+K^-$ and $\gamma K_SK_S$ (with $K_S\to\pi^0\pi^0$) according to PWA results of $\jpsi\to\gamma K_S K_S$~\cite{gammaksks}, as well as $\jpsi\to\gamma \pi^0\eta$, $\gamma\gamma\gamma$, $K_S K_L$, $\pi^0 n\bar{n}$ and $\eta n\bar{n}$ with phase space distribution.
The above MC samples with much larger statistics than in data are helpful to check potential backgrounds.   

\section{DATA ANALYSIS}
\subsection{ANALYSIS METHOD}

In this analysis, the $\jpsi$ sample originates from the decay $\pipijpsi$. The analysis strategy is to first tag $\jpsi$ events by selecting two oppositely charged pions, and then to search for the decay $\jpsigaminv$ within the tagged $\jpsi$ sample.
The branching fraction of the decay $\jpsigaminv$ is calculated using:

\begin{equation}
\label{eq:BR}
\BR = \frac{N_{sig}\cdot\epsilon_{\jpsi}}{N_{\jpsi} \cdot \epsilon_{sig}},
\end{equation}
where $N_{sig}$ and $N_{\jpsi}$ are the yields of the signal candidates of $\jpsigaminv$ and $\pipijpsi$, respectively, and  $\epsilon_{sig}$ and $\epsilon_{\jpsi}$ are the corresponding detection efficiencies, 
evaluated with the corresponding MC samples.
A semi-blind analysis is performed to avoid possible bias, where only one quarter of the full data sample is used to optimize the event selection criteria and to decide upon the upper limit calculation approach. 
The final results are obtained with the full data sample by repeating the analysis only after all the analysis methods are frozen.
In this paper, only the results based on the full data sample are presented.

\subsection{$\jpsi$ TAG PROCEDURE}
$\jpsi$ events are tagged using the two oppositely charged pions produced in the process $\pipijpsi$.
For each charged pion candidate, the point of closest approach to the $\ee$ interaction point must be within $\pm$10~cm
in the beam direction and 1~cm in the plane perpendicular to the beam, and the polar angle $\theta$ with respect to the axis of the drift chamber must satisfy the condition $|\textrm{cos}\theta|<0.93$.
The charged pions are identified by combining the information of the flight time measured from TOF and the $dE/dx$ measured in MDC. 
The corresponding likelihood for the pion hypothesis is required to be larger than that of the kaon hypothesis and 0.001.
To suppress pions not from the decay $\pipijpsi$, the momentum of a pion is required to be less than 0.45~GeV/c.
Additionally, to further suppress the background from $\gamma$ conversion occurring in the inner detector,
the angle between the two selected pions ($\theta_1$) is required to satisfy ${\cos}\theta_1<$0.95.
To veto $\gamma\gamma$ fusion events, the polar angle ($\theta_2$) of the total momentum vector of the pion pair 
should fullfill $|{\cos}\theta_2|<$0.95.

\begin{figure}[!htp]
\begin{center}
	\begin{overpic}[width=0.4\textwidth]{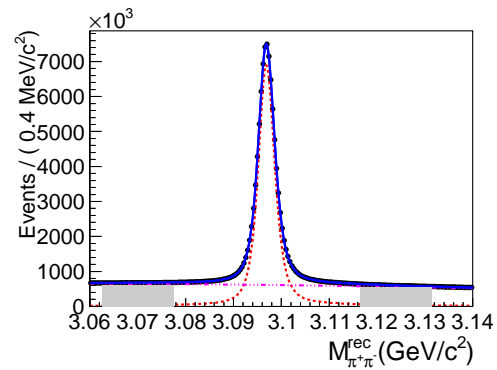}\end{overpic}
	\caption{
	Fit to the $\Mrecpipi$ distribution. The blue solid line is the sum of signal (red dashed line) and background (pink dashed line). 
	The shaded region, (3.0625,3.0775) and (3.1165,3.1315)$\gevcc$, is determined as sideband region for non-$\jpsi$ background study.
	}
	\label{fig:Mpipi}
\end{center}
\end{figure}

To identify $\pipijpsi$ candidate events, the recoiling mass of the $\pp$ system, $\Mrecpipi=\sqrt{(E_{\rm CMS}-E_{\pp})^2-\vec{p}_{\pp}^2}$, is used, where $E_{\rm CMS}$ is the center-of-mass energy of the initial $e^+e^-$ system, and $E_{\pp}$ and $\vec{p}_{\pp}$ are the sum of the energies and momenta of the pions in the rest frame of the initial $\ee$ system, respectively.
The distribution of $\Mrecpipi$ in the range [3.06, 3.14]~$\gevcc$ is shown in Fig.~\ref{fig:Mpipi}, where multiple entries per event are allowed. 
A clear $\jpsi$ peak with low level of background events is observed.
To extract the signal yield, a binned maximum likelihood fit to the $\Mrecpipi$ distribution is performed.
To better model the $\jpsi$ signal shape, a control sample of $\pipijpsi$ with the subsequent decay $\jpsi\to\ee$, which has almost no background, is selected.
In the fit, the signal shape is modeled using the $\Mrecpipi$ distribution of the control sample convoluted with a Gaussian function, which represents the resolution difference  between $\jpsi\to\ee$ and the $\jpsi$ inclusive decay.
The background is described by a 2$^\textrm{nd}$ order Chebychev polynomial function.
The fit results are shown in Fig.~\ref{fig:Mpipi}, and the resolution difference of the $\Mrecpipi$ distribution  between $\jpsi\to\ee$  and the inclusive decay is found to be small, $i.e.$, the width of the Gaussian function is close to zero.
Candidate events in the $\jpsi$ signal region [3.082, 3.112]~$\gevcc$, which is roughly three times the $\Mrecpipi$ resolution, are used for further analysis.
The number of tagged $\jpsi$ events in the signal region is $(8848\pm1)\times10^4$, obtained by integrating the fitted signal curve in the $\jpsi$ signal region.
By performing same procedure on the inclusive MC sample, the efficiency for tagging $\jpsi$ is determined as (56.80$\pm$0.01)\%.

\subsection{SIGNAL SEARCH PROCEDURE}

We search for the decay $\jpsigaminv$ in the remaining $\jpsi$ candidate events by requiring no additional charged track is present and there is exactly one photon candidate.
Photon candidates are reconstructed from EMC and must satisfy the following requirements. 
The minimum energy is ~25 MeV for barrel showers ($|\cos\theta|<$0.80) or 50~MeV for end-cap showers (0.86$<|\cos\theta|<$0.92).
To eliminate showers associated with charged particles, the photon candidates must be separated by at least 20 degrees from any charged tracks in EMC.
To suppress electronic noise or the showers unrelated to the events, the time of the cluster measured from EMC is required to be within 0 and 700 ns after the event start time.
To further suppress background with multiple photons in the final state, the total energy of the remaining showers in the EMC, not satisfying the requirements on photon candidates, is required to be less than 0.1~GeV.
In order to improve the resolution, to further suppress background, and to make sure the invisible particle is within the detector volume, the directions of the signal photon and the missing particle (calculated as the recoiling momentum against the system of $\pp$ pair and signal photon) are required to be within the EMC barrel region.

After the above selection criteria, detailed MC studies indicate that the dominant backgrounds are from $\pipijpsi$ with $\jpsi$ decays into final states including neutral hadrons, $e.g.$, $n\bar{n}$, $\gamma K_L K_L$, $\pi^0 n\bar{n}$.
To further suppress these backgrounds, a series of requirements on the shower shape variables, $i.e.$, 
the second moment should be larger than 5~cm$^2$ and less than 25~cm$^2$, the lateral moment should be larger than 0.1 and less than 0.4, the ratio of energy in $3\times 3$ and $5\times 5$ crystals should be larger than 0.95 due to the narrow shower shape for $\gamma$, 
as well as the number of crystals ($N_{crystals}$) and energy ($E_{shower}$) of the shower should satisfy 4$<N_{crystals}-10\times E_{shower}$ (\gev)$<$20
due to the strong relation between these two variables for $\gamma$, 
are implemented, where these selection criteria are optimized with the control samples of $\gamma$, $\bar{n}/n$ and $K_L$ selected from the decay processes $\jpsi\to\pi^+\pi^-\pi^0~(\pi^0\to\gamma\gamma)$, $\jpsi\to p\pi^-\bar{n}+c.c.$ and $\jpsi\to K\pi K_L$, $\jpsi\to\pi^+\pi^-\phi~(\phi\to K_S K_L)$, respectively.

\begin{figure}[!htp]
\begin{center}
	\begin{overpic}[width=0.4\textwidth]{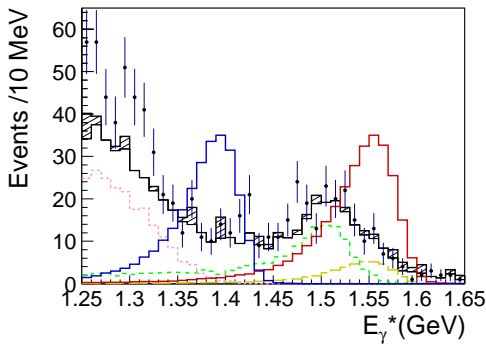}\end{overpic}
	\caption{
	The $\Egam$ distribution. Data is shown with black dots. The total background from $\pipijpsi$, estimated from MC simulation, is shown with the black solid line and includes contributions from the subsequent decays $\jpsi\to\gamma\pi^0$ (long dashed yellow line), $\gamma \eta$ (short dashed green line), and $\gamma K_L K_L$ (dotted pink line). Non-$\jpsi$ backgrounds are estimated using $\jpsi$ sideband events (hatched histogram). The red and blue solid lines show the signal shape with 0 and 1~$\gevcc$ mass assumptions, respectively.}
	\label{fig:Egam}
\end{center}
\end{figure}

The variable $\Egam$, which is defined as the energy of the selected photon in the $\jpsi$ rest frame, is used to identify the signal.
For the signal process $\jpsigaminv$ with a given mass and zero width for the invisible particle, the $\Egam$ is expected to be convoluted with the corresponding detector resolution function.
The distribution of $\Egam$ above 1.25~GeV for the selected events is shown in Fig.~\ref{fig:Egam}.
The dominant backgrounds are from $\pipijpsi$ with subsequent decays $\jpsi\to\gamma K_L K_L$, $\gamma\eta$ and $\gamma\pi^0$, where the latter two produce the peak in the $\Egam$ distribution.
The above three backgrounds, depicted in Fig.~\ref{fig:Egam}, are estimated with the corresponding exclusive MC samples and normalized according to the PDG branching fractions~\cite{PDGweb}.
The contribution from the non-$\jpsi$ process is found to be small and is estimated by the normalized data sample in the $\jpsi$ sideband region (on the  $\Mrecpipi$ distribution), also shown in Fig.~\ref{fig:Egam}.

To better model the peaking backgrounds from $\jpsi\to\gamma\eta$ and $\jpsi\to\gamma\pi^0$ in the follow up procedure,
a binned maximum likelihood fit is performed on the two corresponding exclusive MC samples, individually.
In the fit, the peaking component, where the detected photon is from the $\jpsi$ radiative decay, is described by a Crystal Ball function~\cite{CB}, while the others, which distribute relatively uniformly and correspond to the case that the detected photon is not from the $\jpsi$ radiative decay, is described by a second order Chebychev polynomial function.
The Crystal Ball functions obtained are used to represent the peaking background from $\jpsi\to\gamma\eta/\pi^0$ in the following analysis. 
The number of events are normalized according to the PDG~\cite{PDGweb} and the yield of tagged $\jpsi$ in data.

Unbinned likelihood fits are performed on the $\Egam$ range from 1.25 to 1.65 $\gevcc$,
corresponding to a mass from 0 up to 1.2 $\gevcc$ for the invisible particle.
In the fit, the signal shape is taken from the signal MC simulation convoluted with a Gaussian function representing the resolution difference between data and MC, 
where the parameters of the Gaussian function are obtained by studying a clean control sample of $\pipijpsi, \jpsi\to\gamma\eta~(\eta\to\gamma\gamma)$.
The background shape is described by the sum of an exponential function and two crystal ball functions with fixed amplitudes and shapes presenting for the background of $\pipijpsi$ with subsequent decay $\jpsi\to\gamma\eta$ and $\gamma\pi^0$, respectively, where amplitudes and shapes are estimated by the MC simulation, and the same correction on shape as the signal description is implemented.
~(For heavier invisible particle assumption, the signal shape is broken.)
As no strong peaks are observed in all fits, the upper limits are calculated by using
the modified frequentist method known as $CL_{s}$~\cite{Read:2002hq, Read:2000ru} combined with the asymptotic approximation~\cite{Cowan:2010js}.
In this approach, the test statistic is the profile likelihood ratio, where the likelihood is given with the Possion function:

\begin{equation}
\footnotesize
\label{eq:Likelihood}
\mathcal{L} =  \prod_{i=1}^{N_{bins}} P(N_i|\mathcal{B}\epsilon_{sig}s_iN_{\jpsi}/\epsilon_{\jpsi} + \sum_{j}^{N_{bkg}}b_{ij}^{exp})
\end{equation}
where $s_i$ represents the signal probability in the $i$-th bin,
$\mathcal{B}$ is the branching fraction $\mathcal{B}(\jpsigaminv)$,
$b_{ij}^{exp}$ is the expected background number in the $i$-th bin for the $j$-th source.
Here background is modeled with the exponential function and the two fixed crystal functions from zero-signal assumption fit result.
Additionally, systematic uncertainties are included assuming Gaussian distributions for nuisance parameters.
The upper limit is determined by integrating the test statistic in the range of positive assumed branching fractions.

\subsection{SYSTEMATIC UNCERTAINTIES}

Three categories of systematic uncertainties,  which are associated with the number of tagged $\jpsi$ events~($N_{\jpsi}$), the signal efficiency and the estimated numbers of backgrounds, are considered individually.

The systematic uncertainty related to $N_{\jpsi}$ comes from the binned fit procedure and includes the fit range, bin size, and the shapes of the signal and background.
The uncertainties from the fit range and bin size are estimated to be 0.6\% by varying the fit range by $\pm$ 5~$\mevcc$ and 0.3\% by changing the bin size from 0.4 to 0.2~$\mevcc$, respectively.
The uncertainties from the signal and background shapes are determined as 0.1\%, individually, estimated by the alternative fits without convoluting the Gaussian function on the signal shape or
using a 3-rd order Chebychev function for background.
The total uncertainty related to $N_{\jpsi}$ is 0.7\%, obtained by adding the above components in quadrature. 

To estimate the uncertainty related to the signal efficiency, two control samples, $\ee\to\gamma \ee$ and $\jpsi\to\pi^+\pi^-\pi^0~(\pi^0\to\gamma\gamma)$, are selected.
The former is used to estimate the uncertainty associated with the event topology requirement, {\it i.e.}, no extra photons or charged tracks, as well as the remaining energy requirement.
And the latter is used to estimate the uncertainty associated with the shower shape requirements.
The resulting differences on the efficiency between the data and MC simulation are assigned to be the systematic uncertainty, individually.
The numerical results are 0.6\% and 0.9\% for the ``no extra photons or charged tracks'' requirement and the shower shape requirements, respectively. 
The uncertainty due to the energy cut on the remaining showers in the EMC is less than 0.1\% and negligible.
For the photon reconstruction efficiency, the uncertainty is 1\%~\cite{Ablikim:2010zn}.
By adding all the above uncertainties in quadrature, the systematic uncertainty from the signal efficiency is 1.5\%.

The uncertainties due to the estimated numbers of two peaking backgrounds come from the $\jpsi$ yield,
the decay branching fractions, and the selection efficiency~(or fake rate) for the process $\jpsi\to\gamma\eta/\pi^0$.
The uncertainty of $\jpsi$ yield is discussed above, 0.7\%.
The uncertainties of decay branching fractions are quoted from the PDG~\cite{PDGweb},
3.0\% for $\jpsi\to\gamma\eta$ and 4.8\% for $\jpsi\to\gamma\pi^0$.
The uncertainties associated with the selection efficiency include those of $\gamma$ selection~(including photon reconstruction and shower shape requirements) and the event topology requirement~(including charged tracks number, photon number and extra showers' energy requirements).
The uncertainty associated with the $\gamma$ selection is discussed above.
The uncertainty associated with the event topology requirement is investigated by studying a control sample of $\psip\to\pp\jpsi, \jpsi\to\phi\eta$.
For the decay of $\jpsi\to\gamma \eta$, the control sample is selected by tagging a $\pp$ pair and a $K^+K^-$ pair as well as the $\jpsi$ and $\phi$ mass window requirements on the $\pp$ recoiling system and $K^+K^-$ system, respectively.
The corresponding efficiency is computed for both data and MC samples
by fitting to the $\eta$ signal on the recoiling mass of $\pp K^+K^-$ system 
before and after implementing the event topology requirements.
The resulting difference in the efficiencies is taken as the systematic uncertainty.
For the decay $\jpsi\to\gamma \pi^0$, no extra charged tracks is required, 
since the $\pi^0$ decays into the $\gamma\gamma$ final state dominantly.
Then the same procedure is applied.
Since the efficiency of the event topology requirement is extremely low, $\sim0.2\%/0.3\%$ for the peaking backgrounds of $\jpsi\to\gamma\eta/\pi^0$, the resulting uncertainties, 16\% for both $\jpsi\to\gamma \eta/\pi^0$, are dominated by the statistical uncertainty of the data control sample, and are conservatively taken as the systematic uncertainties in this analysis.
By adding all uncertainties in quadrature, the systematic uncertainties for the number of peaking backgrounds are 17\% for both $\pipijpsi$ and $\jpsi\to\gamma\eta/\pi^0$. 

The uncertainties due to the continuum background, representing by the exponential function, are also included. 
Both the shape and magnitude are considered, and the corresponding uncertainties are evaluated by performing a fit on $\Egam$ distribution with zero-signal assumption.

The all discussed systematic uncertainties are listed in the Tab.~\ref{tab:SysErr}.
\begin{table}[!hbp]
\centering
\caption{Summary of systematic uncertainty }
\footnotesize
\begin{tabular}{ l | c }
\hline
\hline
  source  & uncertainty\\
\hline
\multicolumn{2}{c}{tagged $\jpsi$ number}\\
\hline
signal shape  & 0.1\% \\
\hline
background shape & 0.1\% \\
\hline
fit bin size & 0.3\% \\
\hline
fit range  & 0.6\% \\
\hline
\multicolumn{2}{c}{signal efficiency}\\
\hline
gamma reconstruction&  1\%\\
\hline
only one good shower &  0.6\% \\
\hline
extra showers' energy cut  & less than 0.1\%\\
\hline
shower shape cut &  0.9\% \\
\hline
\multicolumn{2}{c}{fit procedure}\\
\hline
number of $\pipijpsi,\jpsi\to\gamma\eta$&  17\% \\
\hline
number of $\pipijpsi,\jpsi\to\gamma\pi^0$&  17\% \\
\hline
number of continuum background  &  4.4\% \\
\hline
\hline
\end{tabular}
\label{tab:SysErr}
\end{table}

\subsection{UPPER LIMIT RESULT}
\begin{figure}[!htp]
\begin{center}
	\begin{overpic}[width=0.4\textwidth]{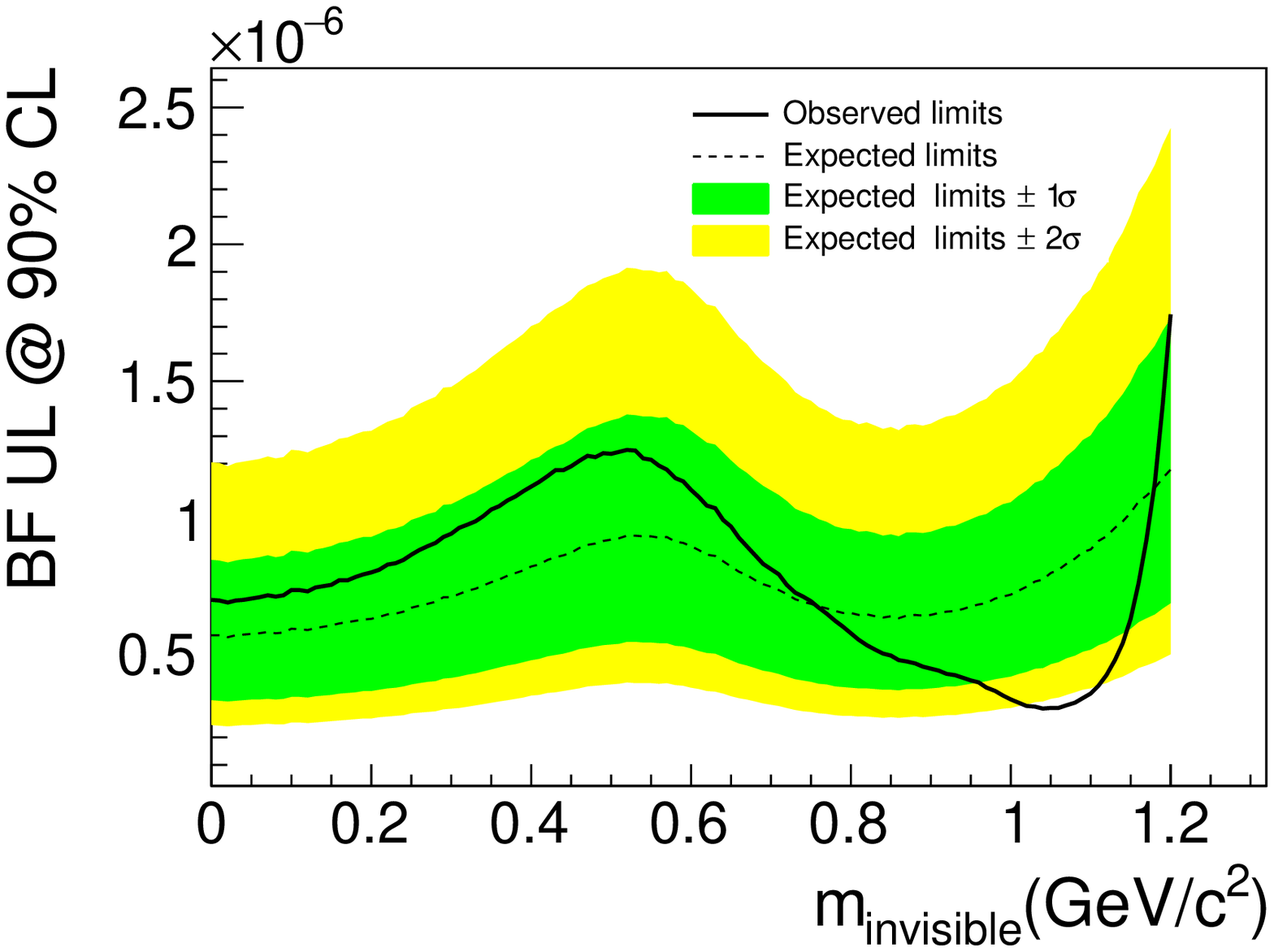}\end{overpic}
    \begin{overpic}[width=0.4\textwidth]{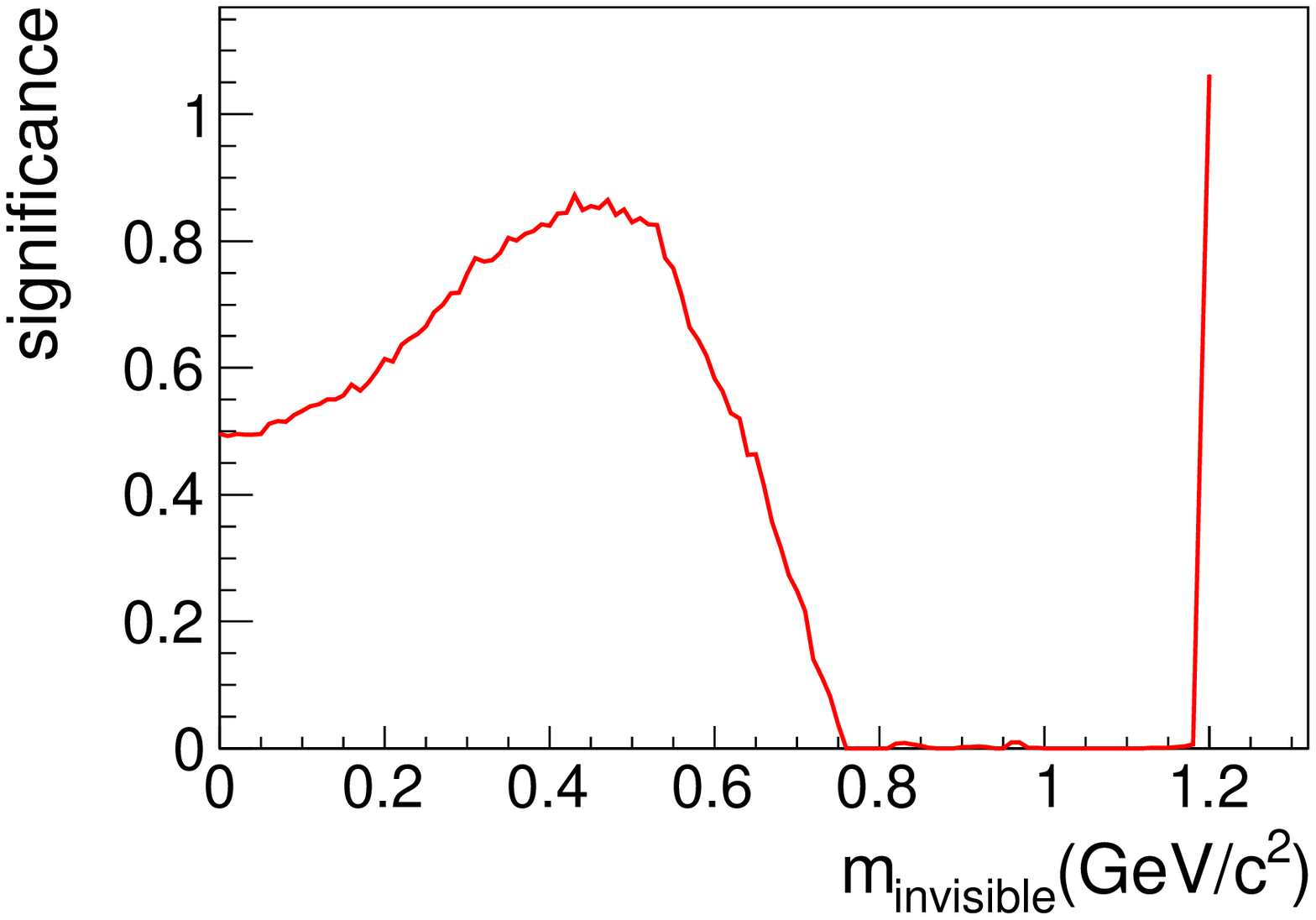}\end{overpic}
	\caption{
Upper limits at the 90\% C.L. for the branching fractions (the upper plot) and the signal significance (the bottom plot) for the decay $\jpsigaminv$.
	In the upper plot, the black line is for data, the black dashed line represents the expected values and the green (yellow) band represents the 1$\sigma$(2$\sigma$) region.
	}
	\label{fig:UL}
\end{center}
\end{figure}

Taking into account all systematic uncertainties and the signal detection efficiencies obtained from MC simulation with different $m_{\rm{invisible}}$ assumptions, the expected upper limits on the branching fraction of $\jpsigaminv$ at the 90\% C.L. are calculated with the $CL_{s}$ approach and are shown in Fig.~\ref{fig:UL}. 
The expected upper limits as well as their uncertainties are also obtained using toy MC sample, which is generated using the background model from no signal assumption fit with the same luminosity as data set.
The result from data is consistent with the zero-signal assumption in the 2$\sigma$ region with most mass assumptions. 
And for the zero mass assumption of the invisible particle the upper limit is 7.0$\times10^{-7}$.
The local signal significances with different mass assumptions are also shown in Fig.~\ref{fig:UL}, where the local signal significance is calculated by $\sqrt{2\mathrm{ln}(\frac{\mathcal{L}_{sig}}{\mathcal{L}_0})}$ incorporating the maximum likelihood with floating signal yield $\mathcal{L}_{sig}$ and with zero-signal yield $\mathcal{L}_{0}$.

\section{SUMMARY and Discussion}

In summary, we search for the $\jpsi$ radiative decay into a weakly interacting neutral particle in the process $\psip\to\pp\jpsi$ by using a $\psip$ sample of $(448.1\pm2.9)\times 10^6$ events collected with the BESIII detector. No significant signal is observed, and the upper limits at the 90\% C.L. on the decay branching fraction of $\jpsigaminv$ are obtained for different $m_{\rm{invisible}}$ assumptions up to 1.2~$\gevcc$. The observed upper limit for a zero mass of the invisible particle is improved by a factor 6.2 compared to the previous results~\cite{Insler:2010jw}.

\begin{figure}[!htp]
\begin{center}
	\begin{overpic}[width=0.4\textwidth]{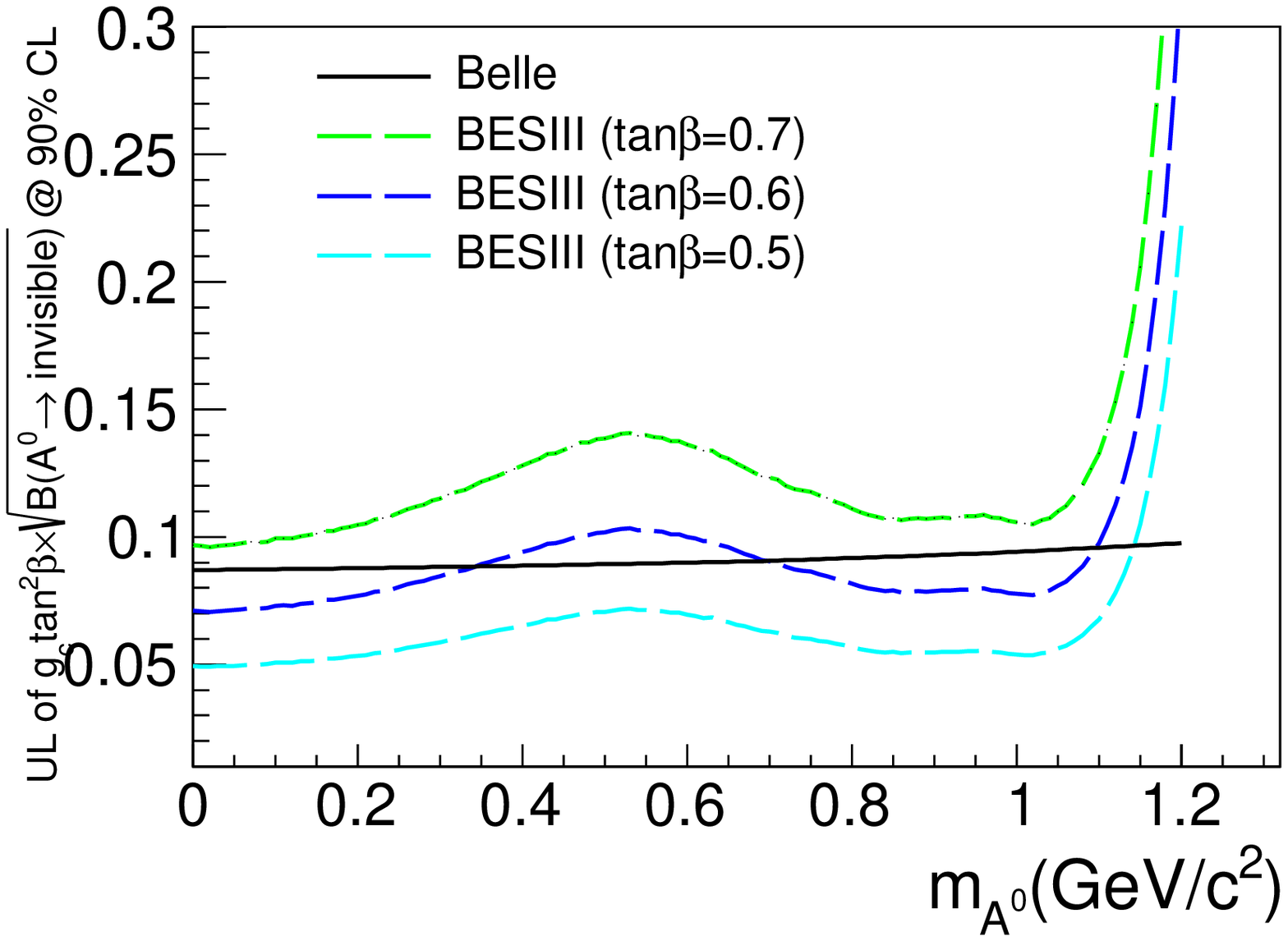}\put(50,3){(a)}\end{overpic}
	\begin{overpic}[width=0.4\textwidth]{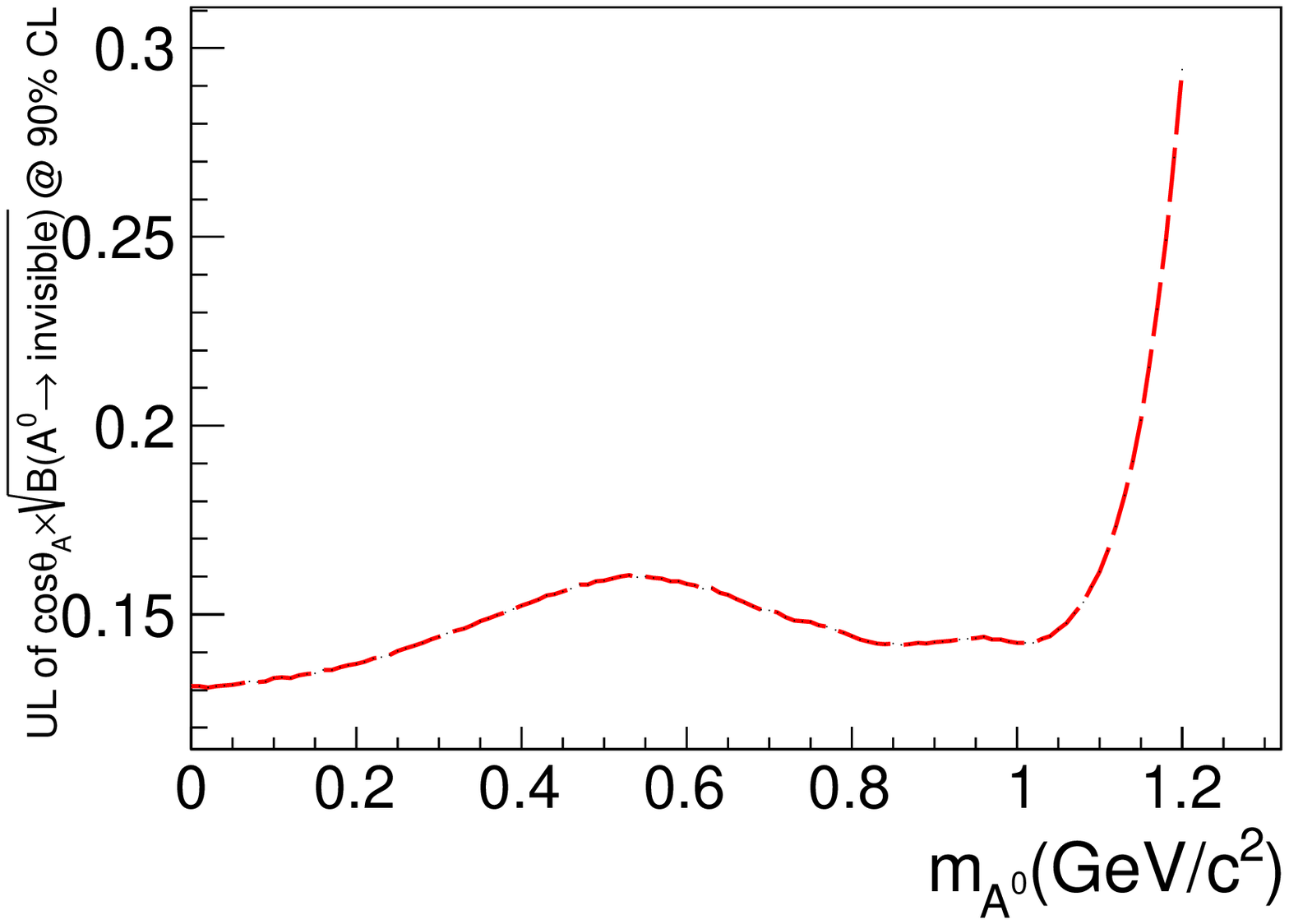}\put(50,3){(b)}\end{overpic}
	\caption[Cal]{
Upper limits at the 90\% C.L. for (a) $g_c\times \tan^2\beta (g_b) \times \sqrt{\mathcal{B}(\rm{A}^0\rightarrow invisible)}$ and
		(b) $\sqrt{g_b g_c} \times \sqrt{\mathcal{B}(A^0\rightarrow invisible)}$.
	}
	\label{fig:Cal}
\end{center}
\end{figure}

To further investigate the physical parameters in NMSSM, and to better compare the physical results from the different quarkonium decays, according to Ref.~\cite{Fayet:2007ua} and Eq.~(\ref{eq:Theory}), the upper limits of $g_c\times \tan^2\beta\times \sqrt{\mathcal{B}(A^0\to invisible)}$ based on the measured upper limits of the $\jpsigaminv$ decay branching fractions are extracted for $\tan\beta=$ 0.7, 0.8 and 0.9, individually, as presented in Fig.~\ref{fig:Cal}~(a).
The extracted results are directly compared to $g_b\times\sqrt{\mathcal{B}(A^0\to invisible)}$($= g_c\times \tan^2\beta\times \sqrt{\mathcal{B}(A^0\to invisible)}$), which is obtained based on the Belle results~\cite{Seong:2018gut} and also presented in Fig.~\ref{fig:Cal}~(a).
We obtain better sensitivity in the range $\tan\beta\le0.6$ compared to the Belle results.
Combining the results from Belle~\cite{Seong:2018gut}, we also extract upper limits on $\cos\theta_A(=\sqrt{g_bg_c}) \times \sqrt{\mathcal{B}(A^0\to invisible)}$, as presented in Fig.~\ref{fig:Cal}~(b).

\section{acknowledgments}
The BESIII collaboration thanks the staff of BEPCII and the IHEP computing center and the supercomputing center of USTC for their strong support. This work is supported in part by National Key Basic Research Program of China under Contract No. 2015CB856700; National Natural Science Foundation of China (NSFC) under Contracts Nos. 11625523, 11635010, 11735014, 11822506, 11835012, 11935015, 11935016, 11935018, 11961141012; the Chinese Academy of Sciences (CAS) Large-Scale Scientific Facility Program; Joint Large-Scale Scientific Facility Funds of the NSFC and CAS under Contracts Nos. U1732263, U1832207; CAS Key Research Program of Frontier Sciences under Contracts Nos. QYZDJ-SSW-SLH003, QYZDJ-SSW-SLH040; 100 Talents Program of CAS; INPAC and Shanghai Key Laboratory for Particle Physics and Cosmology; ERC under Contract No. 758462; German Research Foundation DFG under Contracts Nos. Collaborative Research Center CRC 1044, FOR 2359; Istituto Nazionale di Fisica Nucleare, Italy; Ministry of Development of Turkey under Contract No. DPT2006K-120470; National Science and Technology fund; STFC (United Kingdom); The Knut and Alice Wallenberg Foundation (Sweden) under Contract No. 2016.0157; The Royal Society, UK under Contracts Nos. DH140054, DH160214; The Swedish Research Council; U. S. Department of Energy under Contracts Nos. DE-FG02-05ER41374, DE-SC-0012069.



\end{document}